\documentclass[twocolumn]{aastex63}

\accepted{Dec 17, 2020}
\shorttitle{Patience is a virtue}
\shortauthors{Hansen \& Zuckerman}
\graphicspath{{./}{figures/}}

\begin{document}

\title{ Minimal conditions for survival of technological civilizations in the face of stellar evolution}

\correspondingauthor{B. Hansen; B. Zuckerman}
\email{hansen@astro.ucla.edu, ben@astro.ucla.edu}

\author[0000-0001-7840-3502]{Bradley M. S. Hansen}
\affil{Mani. L. Bhaumik Institute for Theoretical Physics, Department of Physics and Astronomy,
University of California, Los Angeles, CA, 90095; hansen@astro.ucla.edu}

\author[0000-0001-6809-3045]{Ben Zuckerman}
\affil{Department of Physics and Astronomy, University of California, Los Angeles, CA, 90095;ben@astro.ucla.edu}




\begin{abstract}
The ease of interstellar rocket travel is an issue with  implications for the long term fate of our own and other civilizations and for 
the much-debated number of technological civilizations in the Galaxy. We show that the physical barrier to interstellar travel can be greatly reduced if voyagers are patient, and wait for the close passage of another star.   For a representative time of $\sim$1 Gyr, characteristic of the remaining time that Earth will remain habitable, one anticipates a passage of another star within $\sim 1500$~AU. This lowers the travel time for interstellar migration by $\sim$ two orders of magnitude compared with calculated travel times based on distances comparable to average interstellar separations (i.e., $\sim$1 pc) in the solar vicinity. We consider the implications for how long-lived civilizations may respond to stellar evolution, including the case of stars in wide binaries, and the difficulties of identifying systems currently undergoing a relevant close encounter.
 Assuming that life originates only around G-type stars, but migrates primarily to lower mass hosts when the original system becomes uninhabitable, the fraction of extant technological civilizations that exist as diaspora can be comparable to the fraction that still orbit their original host stars.

\end{abstract}



\section{Introduction} \label{sec:intro}

Serious discussion of the existence and prevalence of technological life around other stars dates back, at the least, to 1966 with the appearance of Shklovskii and Sagan's classic book ``Intelligent Life in the Universe'' \citep{SS66}.  
Consideration of these issues naturally leads into the question of whether technological civilizations can migrate from one stellar system to another \citep{SS66,Dy68,Cyclops,RT81,Br82,ZH95,Webb02,Wheeee}.
 The majority of the discussions on this theme focus on the diffusion of settlements away from the origin point under various assumptions regarding the speed at which vessels might travel, and the interval required to launch additional steps. This frequently leaves a model Milky Way teeming with inhabited planets.
  In this paper we wish to examine the opposite limit -- what is the minimum requirement that long term survival of entire civilizations places on interstellar migration? Under this assumption there is no wave of expansion because civilizations would not expand beyond the range needed for survival. We anticipate that this will substantially weaken constraints on the abundance of interstellar civilizations based on the current lack of observed signatures.

Over the past 25 years a variety of observational techniques have enabled rapid advances in the search for, and study of, extrasolar planets. These studies have demonstrated that planetary systems are common around main-sequence stars \citep{How10,Mayor11,CG14,DC15,Burke15,YZ18,HF19,Hsu20}. 
Given the selection effects involved, the vast majority of these planets receive far too much incident radiation to realistically reproduce the kinds of conditions we know to be
amenable to life on Earth, but potentially habitable systems have been discovered at the limits of current technologies 
\citep[e.g.][]{Kep62} and correction for the selection effects suggests   that potentially
habitable planets are quite common \citep{Peti13,DC15,SGW15,HF19,ZH19}.

These discoveries breathe new life into the discussion of the frequency of extraterrestrial technological civilizations. 
If life-bearing planets are common in the Galaxy, and migration between star systems is physically possible, then technological civilizations
could spread rapidly (relative to cosmological timescales) and one would expect their presence to be ubiquitous, including in the Solar system \citep{Hart75}. Yet there is no evidence to suggest they are present.

A commonly suggested solution to this absence of evidence for interstellar migration is to assume that interstellar travel may simply be too difficult or too expensive, and that any civilization that forms on a planet is bound to remain in orbit about that star alone.
 However, there are some well understood physical principles that we know must come into play eventually.
 We know that stars evolve and that Sun-like stars will get larger and more luminous before eventually shrinking and fading away as white dwarfs. Thus, the level of irradiation that a life-bearing planet will experience  undergoes substantial changes in the latter stages of a star's life  and we cannot expect a specific planet to remain habitable forever.
Courses of action are then to either adopt a strategy
of mitigation by moving outwards then inwards while remaining within the gravitational sphere of influence of the host star  \citep[e.g.][]{Gertz} or make the  attempt $-$ however difficult $-$ to migrate to another star \citep{Z85}.  We focus on the latter option here because the mitigation approach requires multiple adjustments to correct for the inescapable evolutionary changes that result in the later stages of the life of the star.

 Thus, if we want to understand the minimum amount of interstellar migration necessary for the long-term survival of a civilization, we must understand how one can minimize the energy expenditure necessary to move from one star to another. In the case of relocation from Earth, the closest star to the Sun, presently, is Proxima Cen, at 4.22 light years distance.
The precise thresholds of feasibility of travelling from Earth to Proxima Cen $-$  in terms the vehicle speed  that can be achieved and the trip duration that can be tolerated $-$ are still a matter of speculation.
Perhaps the only precise statement that can be made on this question is that such a trip is, at present, well beyond our technological and biological 
capabilities \citep[e.g.][]{Ash12}. 
 However, this
  estimate of the time it would take to migrate to the nearest star
is based on an instantaneous snapshot of the relative positions of  
stars in the Solar neighborhood.
Stellar evolution, although inexorable, is also a slow process and impending doom
due to evolution away from the main-sequence
will be readily apparent for a long time beforehand. The key component of our argument is therefore that a civilization under threat will have the opportunity to monitor the positions
of nearby stars and choose the optimal time $-$ when a star passes close by $-$ to attempt a transfer to a different habitat.

 In the following sections, we will estimate the spread of interstellar civilizations under the assumption that they make only enough jumps between stars to satisfy the requirement
of long-term survival.  We will also assume that civilizations only arise in conditions similar to our own, i.e. around G stars. Thus, our model assumes some fraction of G stars produce civilizations capable of sufficient migration to guarantee their survival beyond the end of a main-sequence lifetime, and we wish to understand what requirements this imposes.
 In \S~\ref{Close}  we estimate how close we can expect the Sun (or an equivalent G-type star) to pass near another star, during the course of its main-sequence lifetime. In \S~\ref{Travel}, we estimate how close a star is likely to come during the window of the stellar lifecycle when habitability first becomes threatened. In \S~\ref{Binary} we discuss the consequence when the star is not single but rather a member of a multiple star system. In \S~\ref{Observe} we estimate the requirements to actually find and observe a close encounter wherein a migration might be occurring at the present day, and in \S~\ref{Conc} we summarise our results.

\section{How close an approach can be reasonably expected?}
\label{Close}

The distance of close passage will depend on the local density of stars and the local kinematics,  so the ease of interstellar migration will
depend somewhat on location in the Galaxy. Since our most obvious application is migration from Earth to other stars, or from other stars to Earth,
we begin with the Solar neighborhood.
Let us estimate the local density in two ways.

\cite{Bovy17} measured the local mass density in stars using Gaia DR1, and infers $\rho_* = 0.040 \pm 0.002 \, {\rm M_{\odot}/pc^{3}}$ for the Solar
neighborhood. He measures a mass function for stars with $M > 1 \, {\rm M_{\odot}}$ and extrapolates to lower masses using commonly assumed initial mass functions. To
infer the number density, we use a Kroupa \citep{Kroupa01} mass function to convert this to a number density of $n_* = 0.113 \pm 0.006 \, {\rm pc^{-3}}$
for stars between 0.08 and $8 \, {\rm M_{\odot}}$. We note that there is also a comparable density of brown dwarfs in a Kroupa mass function ($ n_{\rm bd} \sim 0.093\, {\rm pc^{-3}}$)
but we will ignore them as unsuitable locations (they are also poorly constrained by the method above because they make only a small contribution to the total
mass budget).  It is also worth noting that 81\% of the main-sequence stars in this estimate have masses $< 0.5\, {\rm M_{\odot}}$, i.e. we expect the bulk of the
close passages to be with M~dwarfs.  The densityi, $ n = 0.1 \rm pc^{-3}$, that we adopt below implies
an average nearest neighbor distance of 3.85~light years, i.e. the Sun has a very average distance to the nearest neighbor.

A more detailed picture can be obtained by looking at the stellar census of the local 10~pc volume by the REsearch Consortium On Nearby Stars (RECONS) project \citep{Henry18}.
They count 357 total main-sequence stars in this volume, to yield an estimate of $n_* = 0.085 \, {\rm pc^{-3}}$, only about 75\% of the Bovy estimate. 
The greater detail of the RECONS sample is useful because it also identifies which stars are single and which are in multiple systems. If we count independent stellar systems,
then there are 317 (232 single and 85 multiple), which is 89\% of the total stellar count and so  $n_{\rm sys}=0.076 \pm 0.009\, {\rm pc^{-3}}$.   For simplicity, and including systematic errors, we adopt $n_{\rm sys} = 0.10 \pm 0.02\, {\rm pc^{-3}}$ as an average of the two estimates in the calculations that follow.  

As mentioned above, this stellar number density yields an average nearest neighbor distance between stars of 3.85~light years.  However, such estimates rely on the standard snapshot picture of interstellar migration $-$ that a civilization decides to embark instantaneously (at least, in cosmological terms)
and must simply accept the local interstellar geography as is. If one were prepared to wait for the opportune moment, then how much could one reduce the travel distance, and thus the travel time?

To estimate travel times, we need to include Galactic kinematics. \cite{AM20} estimate the velocity dispersions for thin disk stars in the local solar neighborhood as 37~km/s in the radial direction, 24~km/s in the azimuthal direction and 18~km/s in the vertical direction. If we average the three components in quadrature, we find $V \sim 48$~km/s as an approximate three-dimensional
velocity dispersion, which we take to be the average velocity of encounter. 
 Combining this with the above number density, and the concept of a mean free path (= $\tau V$), enables one to calculate a rate of encounter within some distance $R_0$, or an equivalent time $\tau$ between close passages 
\begin{equation}
\tau = 1/(nV \pi R_0^2)
\label{eqn:tau}
\end{equation}
With $n = 0.1 \rm pc^{-3}$ and $V$ = 48~km/s, we can estimate the rate $\Gamma$ at which a given star will encounter others within a distance $R_0$, as\footnote{Gravitational focusing has little effect on these estimates because the focusing factor $\sim G M_{\odot}/R_{\rm 0} V^2 \sim 8 \times 10^{-4}$ even for 
$R_0 = 500$~AU.} 

\begin{equation}
\Gamma = 15.4 \, {\rm Myr^{-1}} \left( \frac{R_0}{1 \rm pc} \right)^2 \label{RateofEncounter}
\end{equation}

 The solar neighborhood is $\sim 8.5$~kpc out from the Galactic center and of only modest stellar density; we expect higher encounter rates in some other parts of the Galaxy. Adopting
the Besan\c{c}on model for the Galactic structure \citep{Besan03}, the density increases as we move through the disk toward the center, peaking at Galactocentric
radii $\sim 2.2$~kpc, with a density $\sim 4.4$ times higher (and similar velocity dispersion). Thus, the stellar encounter rate can get up to
$\Gamma \sim 70\, {\rm Myr}^{-1} (R_0/1 {\rm pc})^2$ in the inner parts of the Galactic disk. Interior to this location, the stellar density of the disk population drops, but the contribution
of the Galactic bulge increases. In the center, the stellar density of the bulge approaches $\sim 14 \, {\rm pc}^{-3}$ \citep{Besan03}, with an increased mean
velocity of 94 km/s. Thus, the encounter rate in the bulge reaches $\Gamma \sim 4 \times 10^{3}\, {\rm Myr}^{-1} (R_0/1 {\rm pc})^2$.

 Even higher stellar densities can be found in the nuclear star cluster at the Galactic center, where \cite{Magoo19} estimates $n \sim 30\, {\rm pc}^{-3}$, with encounter
 velocities $\sim 200$~km/s. This pushes the encounter rate up to $\Gamma \sim 2 \times 10^4 \, {\rm Myr}^{-1} (R_0/1 {\rm pc})^2$. The highest encounter rates are found in globular clusters, where the stellar densities are higher and the velocity dispersions are lower. Characteristic \citep{Harris96} velocity dispersions are $\sim$~10~km/s and half-mass relaxation times are $\sim 10^9$~years, which yield characteristic encounter rates of $\Gamma \sim 10^6\, {\rm Myr}^{-1} (R_0/1 {\rm pc})^2$. In such high encounter rate environments, it is empirically established that dynamical interactions modify not only planetary parameters \citep[e.g.][]{MKJ20} but the properties of the stellar population as well \citep[][and references therein]{BHS13}.

 There are also regions of the Galaxy with lower stellar density than the solar neighborhood. The density of stellar halo stars in the solar neighborhood is $\sim 3 \times 10^{-4}$ of the total, and the averaged three-dimensional dispersion is $\sim 189$~km/s \citep{Besan03}. Away from the disk plane the encounter rate of halo stars with each other is $\Gamma \sim 0.02\, {\rm Myr}^{-1} (R_0/1 {\rm pc})^2$. The encounter rate for a given halo star is dominated by its passage through the Galactic disk, in which case it sees the full surface density of stars from the thin disk, but only for a fraction of its orbital period $\sim 2 \times 2 {\rm kpc}/(2 \pi \times 8.5 {\rm kpc}) \sim 0.08 $ (for solar neighborhood halo stars and a disk half-thickness of 1~kpc) of the time. Incorporating this into the overall estimate yields
$\Gamma \sim 4.5\, {\rm Myr}^{-1} (R_0/ 1 {\rm pc})^2$ for a halo star.


Related work with Gaia data supports the general estimates presented here. Attempts to locate the stars that will pass closest to the Sun in the near future or recent past produce a
close encounter rate \citep{BJR16} within 1~pc  of $\Gamma = 19.7 \pm 2.2 \, {\rm Myr}^{-1}$, in excellent agreement with our estimate in equation~(\ref{RateofEncounter}).  Based on this analysis, the star that is expected to make the closest passage to the Sun in the near future \citep{MO96,BD16,BJR16} is a K7 dwarf Gl~710, which will pass within 13,900~AU of the Sun in approximately $1.28$~Myr. 

 We see that the rates of stellar encounter vary substantially from one Galactic environment to the next, and so the ease of interstellar migration will be a strong function of environment. We will now examine what constraints this places on the energetics of interstellar migration.

\section{Minimum Encounter Distance}
\label{Travel}

  As the estimates in the previous section show, the rate of encounter varies dramatically from one environment to another within the Galaxy. In the highest
density regions -- globular clusters and the nuclear star cluster -- it is well known that the rate of stellar encounters is high enough to substantially
modify the stellar population in observable ways \citep[e.g.][]{FPR75,Bail95,SP95}. This rate of encounters is also sufficient to substantially modify the population
of planetary systems \citep{SS93,DS01,KDS19,WPL20}. In such cases, the issue of long-term survival is likely to be determined by the dynamical evolution of the planetary
system itself, rather than the change in the climate due to stellar evolution. Indeed, it has been suggested \citep{DR16} that globular clusters may offer the
optimally prosperous environment for long-term civilizations.
 However, these high stellar density environments contain only a small fraction of the stellar population of the Galaxy, and are all quite distant from the 
Sun, making observational probes difficult.
Many of these environments are also quite metal-poor, and may have a lower frequency of planetary systems per star.
 The near field environment of the Sun is much less dense, and the influence of neighbors on the dynamical
stability is minimal (after
the dissipation of any potential birth cluster). 

 In the Solar neighborhood, if we multiply the encounter rate of equation~(\ref{RateofEncounter}) by the $\sim 4.6$~Gyr age of the Sun, we find that
we expect the Solar system to have experienced roughly one encounter within 780~AU within that period of time. As the Sun evolves further, it will get brighter and
heat Earth more. By the time the irradiation of the Earth reaches the point at which the greenhouse effect reaches the runaway limit \citep{Ravi13}, the
Sun will have reached an age of $\sim 5.7$~Gyr. Thus, a conservative estimate for the remaining interval of
habitability is $\sim 1$~Gyr (neglecting possible mitigation strategies such as mounting a sunshield at the inner Lagrange point). Within this timeframe, the median distance of closest approach is $\sim 1500$~AU,
 with an 81\% chance that there will be one within 5000~AU. 

 Thus, an attempt to migrate enough of a terrestrial civilization to ensure longevity can be met within the minimum requirement of travel between 1500~and~5000~AU.
This is two orders of magnitude smaller than the current distance to Proxima~Cen.
The duration of an encounter, with the closest approach at 1500~AU, assuming stellar relative velocities of $50 $km/s, is $ 143\, {\rm years}$. In the spirit of minimum
requirements, we note that our current interstellar travel capabilities are represented by the Voyager missions \citep{Voyager}; these, which rely on gravity assists
off the giant planets, have achieved effective terminal velocities of $\sim 20$~km/s. The escape velocity from the surface of Jupiter is $\sim 61$~km/s,
so it is likely one can increase these speeds by a factor of 2 and achieve rendezvous on timescales of order a century.

 This is, of course, a speculative exercise, but the important point to note is that one does not need to postulate significant advances in technological
capabilities to bring interstellar migration within the realms of human possibility. In the time before the Earth becomes uninhabitable, we can expect the Sun
to experience a close enough passage to another star that travel can be achieved on timescales of order of a century with only the gravitational assists from the
giant planets of the Solar system.

 The odds improve further as we move through the Galactic disk closer to the center. As we noted in \S~\ref{Close}, the Besan\c{c}on model has a peak disk density
at $\sim 2.2$~kpc from the center, yielding a higher encounter rate, which reduces the characteristic encounter distance
to $\sim 250$~AU over the course of 10~Gyr (800~AU if we only allow for the last 1~Gyr of the stellar lifetime). 

 Another important component of these rare encounters is the interval between encounters. Above we have estimated the minimum encounter distance within a 
fixed time frame. One could also set a minimum threshold distance for encounters. For instance, if we adopt a threshold distance of 2000~AU, we expect an encounter
within this distance every $\sim 700$~Myr in the solar neighborhood. Thus, even if a civilization attempts to migrate every time a star approaches within this distance,
the opportunity arrives roughly only once  for every few orbits around the Galactic center. Any expansion under these conditions would not be a diffusion away from a central
location, but rather a random seeding. The dispersion in stellar motions would make it difficult to associate any diaspora with its original location.
		
\section{Binary Stars}
\label{Binary}

The discussion in \S~\ref{Travel} is focused on single stars $-$ therefore a civilization that orbits an evolving star is forced to consider migration to another, unbound, star.  But
many stars exist in gravitationally bound multiple systems. This changes the discussion because most binary companions would make a natural, and far easier, destination for migration. Since
our principal interest is those stars which evolve off the main-sequence in a Hubble time, we focus here on the stellar mass range $\sim 0.7$--$1.3 \rm M_{\odot}$. For these
stars, the fraction found in multiple systems is $44 \pm 3 \%$ \citep{Rag10}, so that the calculation in \S~\ref{Travel} applies to approximately half of the G-type stars in the Milky Way. For
the other half,  migration to the environs of the companion (most often an M-type star) would be an option.

The distribution of companion separations is approximately log-normal \citep{DK13,Rag10} and indicates  
 that $\sim$80\% of all G-star binaries will have separations of 1000 AU or less.  Thus, civilizations that arise on solar-like stars in binaries will find it much easier to migrate to their lower mass, longer lived, companion than to a passing star.   However, studies of main-sequence binary systems \citep{WF14} and of binary systems composed of a main-sequence star and a white dwarf \citep{Z14} both suggest that the origin and/or evolution of planets in main-sequence binary systems are disfavored when stellar separations are $<$1000 AU. 
 Planets orbiting members of wide binaries can be destabilized due to the eccentricity excitation of the binary by perturbations from passing stars \citep{KR13}, but these effects occur primarily
 for planets with semi-major axes $>10$~AU,
and so we will assume that this does not affect habitability.
 In summary then, for those binary systems that do spawn technological  civilizations in orbit around the primary, the opportunity to migrate to the binary companion would typically be a far easier option than to another passing star.

 \section{Observing Close Passages}
 \label{Observe}
 
 Attempts to search for evidence of extraterrestrial civilizations are hampered by the very large parameter space that needs to be searched -- see \cite{WKL18} for an attempt to quantify this.
 One interesting consequence of the hypothesis that 
  interstellar migration occurs only during close stellar passages is that it allows one to define a finite set of stars where such events might be occurring at the present time and therefore where possible  evidence of a migration in action might be observed.  Let us consider the density of such targets in the solar neighborhood.
  
  Based on the discussion in \S~\ref{Travel}, let us focus on close encounters with impact parameters  $R_{\rm 0}<2000$~AU. We estimate these occur roughly every 700~Myr for a star in the Solar neighborhood. If we define an encounter time $\sim 2 R_{\rm 0}/V$, this implies that such events last $\sim 400$ years and so any given star spends
  $\sim 400/(7 \times 10^8) \sim 6\times 10^{-7}$ of the time involved in such a close passage. We anticipate that one needs to search $\sim 2\times 10^6$ stars to find one undergoing a close passage at the present day. Given the local stellar density $\sim 0.1\, {\rm pc^{-3}}$, and assuming that $\sim 0.05$ of these are main-sequence G~stars, 
  this implies that we need to search out to a distance $\sim 500$~pc to find a G star undergoing a close passage within 2000~AU at the
  present day.

  
The population of stars in binaries is a substantial foreground contaminant in such a search. As noted in \S~\ref{Binary}, 
$\sim 44 \%$ of the $2 \times 10^6$ G-type stars to be searched will have bound companions. Thankfully, this problem is mitigated by the fact that a substantial fraction of these binaries have separations $<100$~AU \citep{Rag10,DK13}. Using the log-normal period distribution from these references, we estimate that $\sim 10\%$ of G star binaries have separations $>2500$~AU.  Given these considerations, we anticipate $\sim 10^5$ contaminating binaries within a distance of 500~pc. 

Thus, searching for present-day cases of close passages is quite difficult -- removing the binary contamination will require vetting of proper motions and radial velocities.
Fortunately, on scales $\sim 2000$~AU, orbital velocities are far lower than the relative velocities from Galactic kinematics, so any statistically significant velocity discrepancies should be
sufficient to rule out bound pairs. 
 The prospects for investigating such a sample with Gaia may be possible, but nontrivial because of the spatial correlations of proper motion seen on small angular scales \citep{GDR2} and also systematic errors that are introduced for faint stars in the vicinity of brighter ones.
 Thus, extensive vetting of each candidate is required. A preliminary effort to do this is underway but beyond the scope of this paper.

 Searches for evidence of extraterrestrial technology tend to focus either on detecting communication signals \citep{Tarter01,Horo01} or evidence of waste heat in the infrared \citep[e.g.][]{Dyson60,WMS14}. A group of nearby stars undergoing close stellar encounters may provide an alternative set of targets for directed signal searches \citep{KepRad13,Allen16,BTL17,TM17,PMG19}. The large-scale engineering required for substantial interstellar migration could also generate an observable infrared excess, especially if an asteroid population was used to provide raw materials \citep{FM11}. There are also astrophysical reasons to expect increased dust and cometary activity in systems undergoing close encounters, because 5000~AU is approximately the inner edge of the
Solar system Oort cloud \citep{Oort15} and we anticipate that other stars will have similar structures because these scales are set largely by the perturbations of the gravitational field -- see \cite{T93} for a simple scaling explanation. Thus, close stellar passages will penetrate deep into the Oort clouds orbiting other stars and so we can expect them to be sites of enhanced cometary activity anyway. An excess of transient absorptions due to infalling comets -- which have been observed in a handful of stars \citep[e.g.][]{BPic88,Bpic98,WM13,Reb20} -- may prove to be another signature marking a star as worthy of closer scrutiny.

Our results demonstrate that a technological civilization can substantially reduce the physical barrier to interstellar migration by waiting for the opportune moment to transfer to another star.
The capabilities necessary to make such a transfer are still largely a matter of speculation, but patience can reduce the distances required by two orders of magnitude relative to 
the long-term average distance between stars.  If we translate this into a similar reduction in the speed required, it implies four orders of magnitude increase in the amount of transferable mass at fixed energy. Furthermore, travel
at lower speeds is a major advantage  because various
techniques for acceleration and deceleration are available that are not
relevant, or far more difficult, when  applied to relativistic travel.
 Given the extreme energy costs required by interstellar transport, the feasibility of a large-scale transport of mass from one star to another is strongly weighted toward closer encounters.

Thus, even if it is difficult to observe a migration taking place at the present day, the consequences of migration should be reflected in the interstellar density of diaspora today.  In our conservative model -- where migration takes place only in the face of environmental collapse -- there is no diffusion of civilizations away from the original planet, but rather a simple copying process, leaving the original star behind and transferring to a newer, less massive, and more long-lived star.

Therefore, even if life does not start on planets orbiting M dwarfs, some fraction of these systems should be repurposed by transplanted civilizations  and thus should be relevant as a subject for studies of exoplanet habitability \citep[e.g.][]{Tart07}.  In order to estimate the frequency of these diaspora, let us start
by assuming that a fraction $\alpha$ of all G-type main-sequence stars host potentially habitable planets. Only some fraction of these planets may ultimately produce technological civilizations capable of interstellar migration. Thus, we denote $\beta$ as the fraction of all G-type main-sequence stars that yield such civilizations, so that $\beta \leq \alpha$ by definition.
 The density of G stars within 10~pc of the Sun is $\sim 4.5 \times 10^{-3} {\rm pc}^{-3}$\citep{Henry18}. The density of white dwarfs is $5 \times 10^{-3} {\rm pc}^{-3}$. The initial mass function for main-sequence stars above a Solar mass is estimated to have a power law slope $\gamma \sim 2.4$--2.7 (assuming $\rm dN/dM \sim M^{-\gamma}$), based on Gaia DR2 data \citep{Soll19}. This implies that up to 40\% of white dwarfs in the local volume derive from G star progenitors in the range $0.8-1.1 M_{\odot}$ \citep{KHK08,Cumm18}. Thus, the density of white dwarfs in the solar neighborhood that were potentially the point of origin for civilizations that have migrated to another star could be as large as $\sim 2 \times 10^{-3} {\rm pc}^{-3}$.

We previously estimated that $\sim 44 \%$ of G stars are found in binary systems, but also that those with separations less than 1000~AU are likely to experience suppression of planet formation. Based on the log-normal observed distribution of separations, this suggests that $\sim 20\%$ of the binaries featuring G stars may also be viable sites for the birth of technological civilizations. Incorporating these numbers we find that the densities of different classes of targets are 
$\sim 0.56 \times 4.5 \beta \times 10^{-3} {\rm pc}^{-3} \sim 2.5\beta \times 10^{-3} {\rm pc}^{-3}$ for single G stars; $\sim 0.56 \times 2 \beta \times 10^{-3} {\rm pc}^{-3} \sim 1.1 \beta \times 10^{-3} {\rm pc}^{-3}$ for single white dwarfs resulting from G stars; $0.2 \times 0.44 \times 4.5 \beta \times 10^{-3} {\rm pc}^{-3} \sim 4 \beta \times 10^{-4} {\rm pc}^{-3}$ for G stars in binaries and $\sim 0.2 \times 0.44 \times 2\beta \times 10^{-3} {\rm pc}^{-3} \sim 1.8 \beta \times 10^{-4} {\rm pc}^{-3}$ for white dwarfs in binaries that originally evolved from G stars. In total, these represent a density of $\sim 4.2 \beta \times 10^{-3} {\rm pc}^{-3}$ of potentially technologically inhabited systems. In the event that each civilization -- born around a G star
that becomes a white dwarf -- transplants to a longer-lived K or M dwarf, this would imply 30\% of extant civilizations orbit lower mass stars.
Of these low-mass stars, 4\% are in a binary with the parental white dwarf and 26\% are uncorrelated with the original host star.

This estimate does not take into account that a civilization may take a long time to reach the technological sophistication necessary to be observable and/or capable of migration. If we use our own planet as an example, only $\sim 1/5.7 \sim 0.18$ of the possible migration window is spent as a technological civilization. Thus, if we assume only this fraction of the G star samples above host technological civilizations, then accounting becomes
$\sim 4.5 \beta \times 10^{-4} {\rm pc}^{-3}$ for single G stars and $3.2 \beta \times 10^{-5} {\rm pc}^{-3}$ for G stars in binaries. The systems that spawned white dwarfs are, by definition, older and so their densities are unaffected. In this case, $\sim 72$\% of extant civilizations are associated with lower mass stars (10\% in white dwarf binaries and 62\% around stars outside the original birth system.)

\section{Conclusions}
\label{Conc}

 Our principal focus is to estimate the barrier faced by a technologically advanced civilization in the face of the late-stage (red giant to white dwarf) evolution of the host star. The question is whether it is easier to accommodate large changes in stellar luminosity and mass, or to transfer some or all of the civilization to another star.
 We demonstrate that transfer of a civilization to another star system can be accomplished at modest velocities in a time comparable to a current human lifetime provided that the migrants wait for close passage of another star.  In the solar vicinity one expects appropriate close passages $-$ $\sim$100 times smaller than typical stellar separations $-$ to occur at least once during a characteristic time of a Gyr.   Most encounters will be with long-lived K- or M-type stars on the
main-sequence, and so few civilizations will be required to make more than one transfer in a Hubble time.

 Our hypothesis  bears resemblance to the slow limit in models of interstellar expansion \citep{WMS14,CN19}. In a model in which civilizations diffuse away from their original locations with a range of possible speeds, the behavior at low speeds is no longer a diffusion wave but rather a random seeding dominated by the interstellar dispersion. Even in this limit, the large age of the Galaxy allows for widespread colonization unless the migration speeds are sufficiently small. In this sense our treatment converges with prior work, but our focus is very different. We are primarily interested in how a long-lived technological civilization may respond to stellar evolution and not how such civilizations may pursue expansion as a goal in and of itself. Thus our discussion demonstrates the requirements for technological civilizations to survive the evolution of their host star, even in the
event that widespread colonization is physically infeasible.

 We wish to estimate the observational consequences of this minimal model.
 To that end, we provide three observational estimates.

 The first is simple enough. If survival is the sole goal, then a civilization need only transfer once  to a long-lived
K or M dwarf  to ensure its survival for a Hubble time. In that event, the average density of technological planetary systems will reflect the frequency with which they
arise, except that the host stars will be biased toward lower masses even if life only starts on G-type stars. In \S~\ref{Observe} we
estimated that the local density of M stars -- both single and in binaries -- hosting transplanted civilizations is $\sim 1.3 \beta \times 10^{-3} {\rm pc^{-3}}$, where $\beta$ is the fraction of G stars that host civilizations capable of interstellar migration.
 The corresponding density of systems still in orbit about
the original G star  will depend on how quickly technological civilizations arise. If this happens quickly, then systems orbiting their original G stars will still outnumber diaspora. However, if it takes 4.5~Gyr for a civilization to become observable, then diaspora may outnumber the observable G star systems. In this case, and assuming $\beta \sim \alpha \sim 0.3$ -- an optimistic interpretation of current estimates of the frequency $\alpha$ of habitable planets around G stars \citep{ZH19,Bry20} -- we would anticipate two systems within 10~pc of the Sun to host technological civilizations, and at least one of them to orbit a low-mass K or M star.

 A second consequence of our hypothesis is that civilizations that form around G stars in binaries have the easiest route to long-term survival, because the transfer to a lower mass binary companion is much easier than migration to a gravitationally unbound star. M and K dwarfs orbited by white dwarfs are thus likely to be a fruitful sample to search for civilizations advanced enough to have achieved the feat of transferring from one star to another.
In \S~\ref{Binary} we estimated the density of such systems to be  $ 1.8 \beta \times 10^{-4} {\rm pc}^{-3}$. Although this is lower than the density of unbound diaspora, the energy requirements for migration are much less and so more reliable.

 The third consequence is that we associate the migration with a particular astrophysical event that is, in principle, observable, namely a close passage of two
stars. One could reduce the vast parameter space of a search for evidence of technology with a focus on such a sample of stars in a search for communication signals or signs of activity such as infrared excesses or transient absorptions of stellar photospheres. However, our estimates suggest that the density of such systems is low compared to the confusing foreground of truly bound stars, and a substantial program of vetting false positives would be required.


\acknowledgements We thank Dr. Beth Klein for assistance.  This research was supported in part by grants to UCLA from NASA and NSF. This research has made use of NASA's Astrophysics Data System and the results of the RECONS consortium, which can be found at  www.recons.org. We thank the referee for a timely and enthusiastic correspondence.

%




\bibliography{refs.bib}{}
\bibliographystyle{aasjournal}



\end{document}